# Decoupling the Device and Identity in Cellular Networks with vSIM

Shirin Ebadi
Zach Moolman
University of Colorado Boulder
shirin.ebadi@colorado.edu
zach.moolman@colorado.edu

Eric Keller
Tamara Lehman
University of Colorado Boulder
eric.keller@colorado.edu
tamara.lehman@colorado.edu

## I. INTRODUCTION

Cellular networks are now fundamental infrastructure, powering not just smartphones for daily communication and commerce, but also enabling the expansion of IoT and edge computing through last-mile connectivity. At the core of this infrastructure is the SIM card, which provides essential network authentication and subscriber identification through subscriber cryptographic key and profile information. More recently, the SIM card has evolved from a separate pluggable card, to a card integrated into the board (i.e., soldered onto the board with the same electrical interface) (eSIM), to one that is integrated into the System on Chip (iSIM).

However, a fundamental limitation persists across SIM evolution: subscriber identity remains coupled to hardware. eSIM and iSIM technologies, despite enabling remote provisioning, still bind digital identities to specific hardware elements. This makes it complex to support emerging use cases like moving a phone number to a cloud AI service or transferring credentials between different devices while maintaining cellular connectivity. Furthermore, although eSIM and iSIM support multiple profiles (multiple phone numbers or carrier profiles on a single device), all profiles still link back to the same hardware identity. For users seeking to maintain privacy through identity rotation or separation (like having different numbers for different purposes), they are limited by the hardware-bound nature of the security architecture.

In this paper, we seek to decouple identity from the device, enhancing privacy and flexibility compared to various SIM designs. By breaking this coupling, we enable scenarios like real identity rotation, integration with virtual assistants, or temporary use of backup phones while maintaining consistent cellular connectivity.

## II. vSIM

Modern SIM security fundamentally relies on hardware binding for authentication and trust. eSIM and iSIM rely on manufacturing-time ID, certificates, and keys installed in their secure hardware elements for profile provisioning and management. These credentials, used to establish secure channels with profile management servers, create a permanent binding between the subscriber profile and the hardware identity. To break this binding, we introduce vSIM, a software-based SIM implementation that operates within Trusted Execution Environments (TEEs) and enables secure digital provisioning. Instead of using hardcoded IDs and keys for identity verification, vSIM proves the execution of specific trusted software. This is achieved by leveraging Enhanced Privacy ID (EPID) [1], where each CPU maintains a unique private key while sharing a group public key. This approach enables anonymous attestation—devices can prove their authenticity to carriers without revealing individual identities—while supporting fine-grained security management through multiple revocation mechanisms (e.g., signature pattern blacklisting). During profile provisioning, vSIM signs its attestation quote using its EPID private key, allowing providers to verify authenticity and trustworthiness of vSIM against group keys and revocation lists before establishing a secure channel for profile delivery.

### A. Design Overview

Figure 1 illustrates the high-level architecture of vSIM. (1) vSIM executes on the device's main CPU within a TEE, providing hardware-enforced isolation between vSIM's secure execution and the normal operating system. It implements standard SIM card functions for 5G authentication and adds secure provisioning capabilities. (2) The trusted hardware includes EPID private key which is used for authentication. (3) The remote attestation measures the vSIM binary and signs it using EPID private key. (4) The provider vSIM manager, residing on the provider's infrastructure, handles profile management and attestation verification for vSIMs requesting new profiles. (5) After successfully downloading a profile, vSIM will store it permanently in a secure storage.

### B. Establishing Trust and Secret Key Provisioning

vSIM is initially installed without a subscriber profile and must securely obtain one from a user-specified provisioner.

**Trust Establishment.** As a software solution, vSIM builds trust through secure boot and remote attestation. Secure boot provides the foundation by creating a chain of trust where each boot layer verifies the next layer's signature, ultimately ensuring the TEE and attestation system boot securely. The attestation system, running within this verified TEE environment, then generates cryptographically signed quotes that prove vSIM's integrity and authenticity to remote parties.

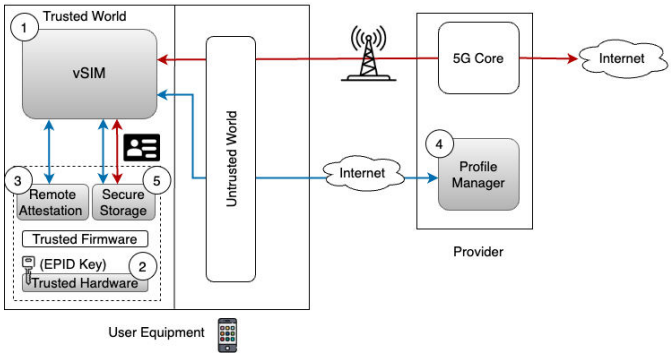

Fig. 1: vSIM Architecture (Blue arrow shows profile provisioning and red arrow represents authentication procedure).

**Secure Channel Protocol.** The provisioning process begins when vSIM generates a nonce and ephemeral public key, then transmits them encrypted with the provisioner's public key. The provisioner transmits back nonces, an attestation request, and its ephemeral public key, encrypted with vSIM's public key. Both parties derive a session key from their ephemeral keys. vSIM then sends its attestation quote (signed by the private key) and nonces, encrypted with the session key. Upon verifying the quote and signature by the provisioner, the secure channel is established. The provisioner then securely transfers the subscriber profile, which includes both the user's subscription data and the secret key needed for 5G authentication. This protocol ensures mutual authentication between vSIM and provisioner, protection against replay attacks through nonces, verification of vSIM's trusted execution state, and forward secrecy through session keys.

*C. Using vSIM in 5G*

After obtaining the subscriber profile and the secret key, we need to securely persist them in storage. In order to minimize the software running in the TEE, the data is encrypted and then passed to the file system running in the untrusted OS.

Finally, vSIM provides all essential functions to support 5G authentication. The process follows the traditional flow: when the Mobile Equipment (ME) receives a network challenge, it forwards it to vSIM. The vSIM retrieves and decrypts the secret key from storage, generates the required cryptographic keys and response, and securely transmits them back to the ME for network verification (Figure 1 red arrow).

## III. Preliminary Work

We implemented vSIM in the Keystone [2], an open-source TEE framework for RISC-V, deploying it in a Qemu-emulated environment. For cryptographic functions, we ported Libsodium library along with tiny AES libraries into the enclave runtime. In accordance with specifications in 3GPP TS 33.501 standard [3], we implemented 5G authentication using AES-CBC with 256-bit keys. We provided a remote attestation program which resides within Keystone and provides vSIM's binary measurement and its signature when requested.

For profile provisioning, we developed a Python-based vSIM Manager server that handles profile provisioning and attestation verification.

## IV. Integration with 5G structure

To verify that vSIM seamlessly integrates with 5G infrastructure, we leveraged srsRAN with ZeroMQ, extending srsUE's source code to interface with vSIM instead of USIM.

We measured the traffic with/without vSIM. The 5G authentication process is done only once when the UE wants to attach to the network, and therefore communication can occur without vSIM's involvement afterwards. Using iPerf3, we measured TCP traffic performance by sending 200 MB through the UE. Figure 4(a) illustrates network usage rates using vSIM, and Figure 4(b) shows it with USIM. Apart from some noise, they are roughly identical, illustrating the fact that there is no overhead.

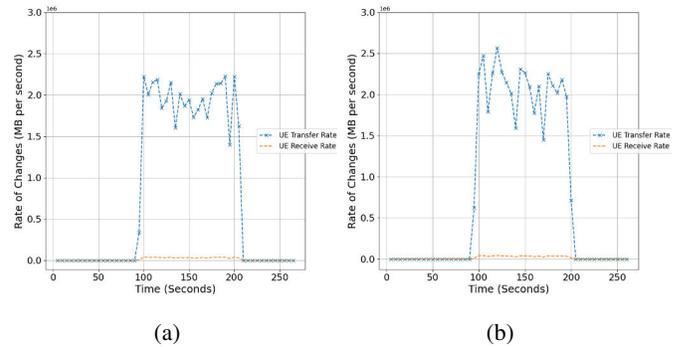

Fig. 2: Inbound and Outbound network usage of UE. (a) With vSIM, (b) With USIM

## V. Conclusions and Future Works

In this paper, we introduce vSIM, a software-based reimagining of SIM functionality that addresses limitations of traditional SIM cards. Our implementation integrates with commercial off-the-shelf CU/DU solutions like srsRAN, validating vSIM's compatibility with existing network infrastructure.

However, our work here is only the beginning. As part of our future work, we plan to integrate vSIM into an FPGA implementation of Keystone for IoT devices. With this, we aim to implement discussed features such as multi-profile support which enables comprehensive evaluations of performance, compatibility, and cost benefits compared to traditional SIM solutions in real-world deployments.

## References

[1] E. Brickell and J. Li, "Enhanced privacy id: A direct anonymous attestation scheme with enhanced revocation capabilities," *IEEE Transactions on Dependable and Secure Computing*, vol. 9, no. 3, pp. 345–360, 2012.
[2] D. Lee, D. Kohlbrenner, S. Shinde, K. Asanović, and D. Song, "Keystone: an open framework for architecting trusted execution environments," in *Proceedings of the Fifteenth European Conference on Computer Systems*, ser. EuroSys '20. New York, NY, USA: Association for Computing Machinery, 2020. [Online]. Available: https://doi.org/10.1145/3342195.3387532
[3] "3GPP TS 33.501 version 16.3.0 Release 16," 3rd Generation Partnership Project (3GPP), Tech. Rep., 2020, accessed: 2025-01-01.